\documentclass[
aps,
longbibliography,
superscriptaddress,
 amsmath,amssymb,
twocolumn,
]{revtex4-1}

\usepackage{graphicx}
\usepackage{dcolumn}
\usepackage{bm}
\usepackage{upgreek}
\usepackage{color}
\usepackage{hyperref}
\usepackage{amsmath}

\newcommand{\m}{\mathrm}

\newcommand{\fref}[1]{Fig.~\ref{#1}}

\definecolor{gold}{RGB}{215,155,0}
\definecolor{blue}{RGB}{0,0,255}
\definecolor{red}{RGB}{255,0,0}
\definecolor{darkgreen}{RGB}{20,150,10}
\definecolor{darkblue}{RGB}{10,10,150}
\definecolor{orange}{RGB}{200,100,0}
\definecolor{lightblue}{RGB}{50,150,230}

\usepackage[normalem]{ulem}

\begin{document}

\title{Coupling high-overtone bulk acoustic wave resonators via superconducting qubits}

\author{Wayne Crump}
\affiliation{Department of Applied Physics, Aalto University, P.O. Box 15100, FI-00076 AALTO, Finland}%

\author{Alpo V\"alimaa}
\affiliation{Department of Applied Physics, Aalto University, P.O. Box 15100, FI-00076 AALTO, Finland}%




\author{Mika A. Sillanp\"a\"a}
\affiliation{Department of Applied Physics, Aalto University, P.O. Box 15100, FI-00076 AALTO, Finland}%


\begin{abstract}
In this work, we present a device consisting of two coupled transmon qubits, each of which are coupled to an independent high-overtone bulk acoustic wave resonator (HBAR). Both HBAR resonators support a plethora of acoustic modes, which can couple to the qubit near resonantly. We first show qubit-qubit interaction in the multimode system, and finally quantum state transfer where an excitation is swapped from an HBAR mode of one qubit, to an HBAR mode of the other qubit.
\end{abstract}

\maketitle

%

Hybrid quantum systems seek to combine strengths and offset weaknesses of different quantum technologies in order to improve capability beyond that of any one technology. Superconducting circuits are one of the more mature quantum technologies at this stage and have been integrated with many other systems due to the relative ease in design and fabrication as well as good coherence times \cite{Clerk2020hybrid}. 

Many different acoustic systems have been integrated with superconducting circuits such as nanomechanical oscillators \cite{Lehnert2008Nph,Teufel2011a,OConnell2010}, phononic crystals \cite{Safavi2019Fock}, bulk acoustic wave systems \cite{SchoelkopfHBAR2017,kervinen_interfacing_2018} and surface acoustic wave systems \cite{Delsing2014,Nakamura2017,moores_cavity_2018,Cleland2019PhEntangl}. Acoustic resonators can offer great coherence properties \cite{gokhale_epitaxial_2020} as well as smaller mode volumes due to the relation between wave velocity and wavelength, with the difficulty coming in coupling these resonators strongly with electromagnetic systems. 

The strong coupling of acoustic modes with superconducting qubits has resulted in many experiments exploring the quantum nature of mechanical oscillations, with experiments demonstrating number splitting \cite{Safavi2019Fock}, the creation of non-classical states in the acoustic mode \cite{chu_creation_2018}, Landau-Zener-Stückelberg interference \cite{kervinen2019landau}, and entanglement \cite{Wollack2022entangle}.  The ability to prepare acoustic resonators in arbitrary quantum states opens up the possibility of using them in applications such as quantum memories due to their coherence properties and insensitivity to electromagnetic noise.

High-overtone bulk acoustic wave resonators (HBAR) offer access to mechanical modes in the GHz regime, making them attractive for integration with superconducting qubits. The piezoelectric interaction enables coupling in the strong regime and their state to be controlled and read-out using the qubit. The system has been implemented using a 3D \cite{SchoelkopfHBAR2017} and 2D \cite{kervinen_interfacing_2018} transmon architecture with part or all of the qubit capacitor directly patterned on the piezo layer of the HBAR. This was later improved in both cases by using a flip-chip design \cite{chu_creation_2018,Kervinen2020} which has lead to the current state of the art \cite{Lupke2022}. Experiments on these system have demonstrated the creation of non-classical multiphonon states \cite{chu_creation_2018}, demonstration of dispersive readout for a parity measurement of the mechanical mode \cite{Lupke2022}, and sideband control of the mechanical modes \cite{Kervinen2020}.

Work thus far has focused on coupling of a qubit and a single HBAR device supporting a set of acoustic modes. In this work we couple two complete qubit-HBAR systems together via qubit-qubit interaction, and transfer excitations within the system, including between the HBAR modes. This demonstrates the possibility of integrating multiple HBAR devices into quantum circuits enabling the exploration of much larger and complex systems.

In the system there are two qubits which are coupled together as well as being individually coupled to a set of HBAR modes. The qubit-mode couplings can be described by the Jaynes-Cummings model, and the qubit-qubit coupling will be capacitive and therefore expected to take the iSWAP form
\cite{Kwon2021GateBased}. The system as a whole can then be described by the Hamiltonian:
\begin{equation}\label{eq:SystemHamiltonian}
    \begin{split}
        H/\hbar = & \frac{\omega_1}{2}\sigma_{(z,1)} + \frac{\omega_2}{2}\sigma_{(z,2)} + J \left(\sigma_{(+,1)} \sigma_{(-,2)}  + \sigma_{(-,1)} \sigma_{(+,2)} \right) \\
         + & \sum_m \left[ \omega_{(m,1)} \left( a_{(m,1)}^\dagger a_{(m,1)} + \frac{1}{2}\right)  \right. \\
         & \left. + g_{(m,1)} \left(a_{(m,1)}^\dagger \sigma_{(-,1)} + a_{(m,1)} \sigma_{(+,1)}\right)\right] \\
         + & \sum_n \left[ \omega_{(n,2)} \left( a_{(n,2)}^\dagger a_{(n,2)} + \frac{1}{2}\right) \right. \\
         & \left. + g_{(n,2)} \left(a_{(n,2)}^\dagger \sigma_{(-,1)} + a_{(n,2)} \sigma_{(+,1)} \right)\right] \,,
    \end{split}
\end{equation} 
where $\omega_1$ and $\omega_2$ are the qubit frequencies, $J$ is the qubit-qubit coupling, $\omega_{(m,1)}$ and $\omega_{(n,2)}$ are the HBAR mode frequencies corresponding to their respective qubits and $g_{(m,1)}$, $g_{(n,2)}$ are the couplings to the HBAR modes. The $\sigma_{i,j}$ are the pauli operators and $a_m,a_m^\dagger$ are the annihilation and creation operators.

In order to theoretically analyze the experiments described below, we determine the time evolution of the system using the Lindblad master equation. We include the qubits' decay and dephasing, as well as mechanical mode decay.




\begin{figure}[h!]
\centering
\includegraphics[width=0.95\columnwidth]{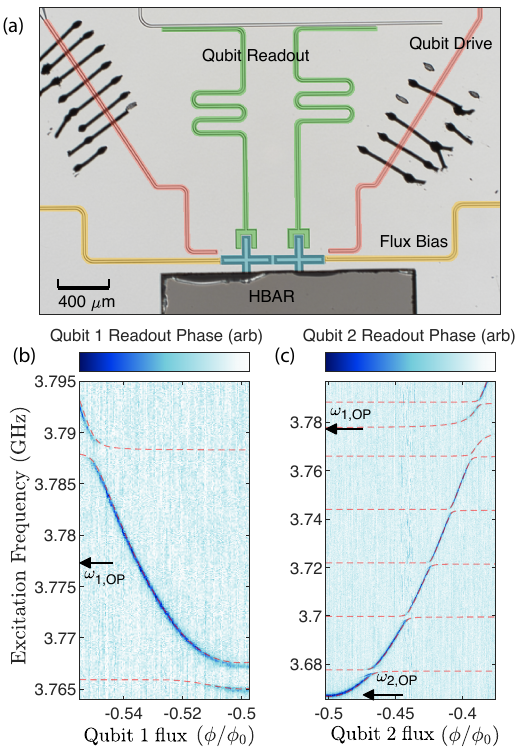}
\caption{\emph{Sample Overview:}  {\bf(a)} Optical image of the sample used in the experiment. It consists of two capacitively coupled qubits (blue) and an HBAR flip chip over the two qubits. Each qubit has separate control lines for flux (yellow) excitation (red), and a readout resonator (green). {\bf(b)} Two tone spectroscopy of qubit 1 around its operating point.  {\bf(c)} Two-tone spectroscpy of qubit 2 around its operating point. In (b-c), the eigenvalues of equation \ref{eq:SystemHamiltonian} are plotted on top as dashed lines, and the operation points of the qubits in the transfer experiments are labeled.}
\label{fig:overview}
\end{figure}

Figure \ref{fig:overview} shows an optical image of the device used for the experiments. 
The device consists of a superconducting circuit with two qubits, each with their own readout, flux bias control and excitation lines. The qubits have a capacitive coupling to each other, as well as to the HBAR flip chip that covers both. The qubits have a round pad on the bottom arm of around 80 $\mu$m in diameter which defines the capacitive coupling to the HBAR chip. The circuit was patterned using Electron beam lithography and metalised with evaporated Aluminium. Double angle evaporation was used to create the Josephson junctions for the qubits. 

The HBAR flip chip consists of a 900 nm AlN piezo layer, a 250 $\mu$m sapphire layer and a 60 nm Mo layer in-between to act as a ground plane to enhance the coupling to the mechanical modes \cite{Kervinen2020}. The HBAR was placed by hand onto the circuit chip and glued with standard epoxy. 

The qubit frequencies can be tuned in the range 3.7-4.5 GHz and have readout resonator frequencies of 6.230 GHz and 6.013 GHz.  The operating points of the qubits were chosen to maximise their coherence properties and hence they are operating at or close to their minimum frequencies, as shown in \fref{fig:overview}.

The bottom two plots of figure \ref{fig:overview} show two-tone measurements sweeping the qubit frequencies in the neighbourhood of their operating frequencies chosen for later experiments. The operating frequency of qubit 1 was set near its minimum at $\omega_{1,\m{OP}}/2\pi = 3.7778$ GHz and qubit 2 at its minimum at $\omega_{2,\m{OP}}/2\pi = 3.6673$ GHz. The many small anticrossings occur when a qubit is sweeping past an HBAR mode, while the larger anticrossing at 3.778 GHz seen in the data for qubit 2 corresponds to the qubit-qubit coupling. The spacing between HBAR modes (free spectral range, FSR) is around 22 MHz which corresponds well with the thickness of the HBAR sapphire layer. The dashed lines show the eigenvalues according to equation \ref{eq:SystemHamiltonian}.

\begin{figure}[h!]
\centering
\includegraphics[width=0.9\columnwidth]{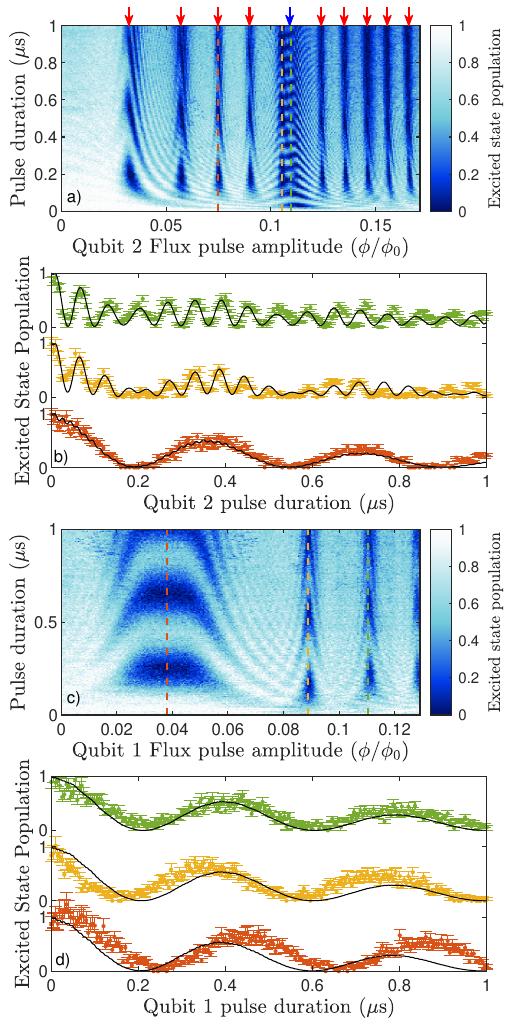}
\caption{\emph{Vacuum Rabi oscillations:} {\bf (a)} Qubit 2 is first excited and then a square pulse of variable length and amplitude is applied to its flux control. Vacuum Rabi oscillations are seen between the right qubit and its coupled HBAR modes (red arrows) as well as with qubit 1 (blue arrows). {\bf (b)} Line cuts at the points indicated by the dashed lines in (a) are plotted as points with lines calculated using the Lindblad master equation and equation \ref{eq:SystemHamiltonian} (black lines). {\bf (c)} The same experiment as before but using qubit 1. Now we see vacuum Rabi oscillations between the qubit 1 and its coupled HBAR modes. {\bf (d)} Line cuts at the points indicated by the dashed lines in (c) are plotted as points with lines calculated using the Lindblad master equation and equation \ref{eq:SystemHamiltonian} (black line).}
\label{fig:NormalVR}
\end{figure}

At the qubits respective operating points, they had $T_1$ values of 2.2 $\mu$s and 2.41 $\mu$s, as well as $T_2$ values of 4.41 $\mu$s and 1.02 $\mu$s. Their respective 2$g$ couplings to their HBAR modes were 2.55 MHz and 2.85 MHz, with the mechanical $T_1$ values being 380 ns and 320 ns. The system had a qubit-qubit 2$g$ coupling of 16.7 MHz.

Figure \ref{fig:NormalVR} shows a vacuum Rabi oscillation experiment where an excitation is swapping between an initially excited qubit and its coupled mechanical modes. In panels (a,b) qubit 2 is being controlled and measured and we see vacuum Rabi oscillations with the mechanical modes (red arrows) and also with the other qubit (blue arrows), corresponding with the anticrossings seen in figure \ref{fig:overview} bottom right. In figure \ref{fig:NormalVR} (c,d) qubit 1 is controlled and experiences vacuum Rabi oscillations with its coupled mechanical modes following the anticrossings seen in figure \ref{fig:overview} bottom left. Since the flux is tuned in the positive direction, it first sweeps on resonance with the lower mode and then with the upper mode seen in figure \ref{fig:overview} bottom right.


If one looks closely the vacuum Rabi oscillation fringes can be seen to be asymmetric, especially in figure \ref{fig:NormalVR} (a). The source of this unknown and it results in deviations from the theory at later simulation times. Some slight asymmetry could be generated for the nearest mode by including the effect of the $\pi$ pulse specifically in the simulations, but this was not enough to reproduce the long tail of the fringes from the mode nearest the qubit operation point seen in figure \ref{fig:NormalVR} (a) which extend very far, up to where qubit 1 is. It can also be seen in figure \ref{fig:NormalVR} (a) that the vacuum Rabis with qubit 1 also show these extended fringes on the right side. This behaviour may be related to the same phenomena that is seen in frequency domain, where at the avoided crossing, the upper branch has less weight than the lower branch. It is possible at least some of the asymmetry is caused by pulse distortion \cite{Rol2020PulseDistorsion}.

The line cuts in figure \ref{fig:NormalVR} (b) show a double oscillation feature that occurs when qubit 2 is near the qubit 1 frequency. This is because the excitation is experiencing Rabi oscillations with both the other qubit and the nearby acoustic modes at the same time but on different time scales, hence the multiple oscillating feature.

\begin{figure}[h!]
\centering
\includegraphics[width=0.9\columnwidth]{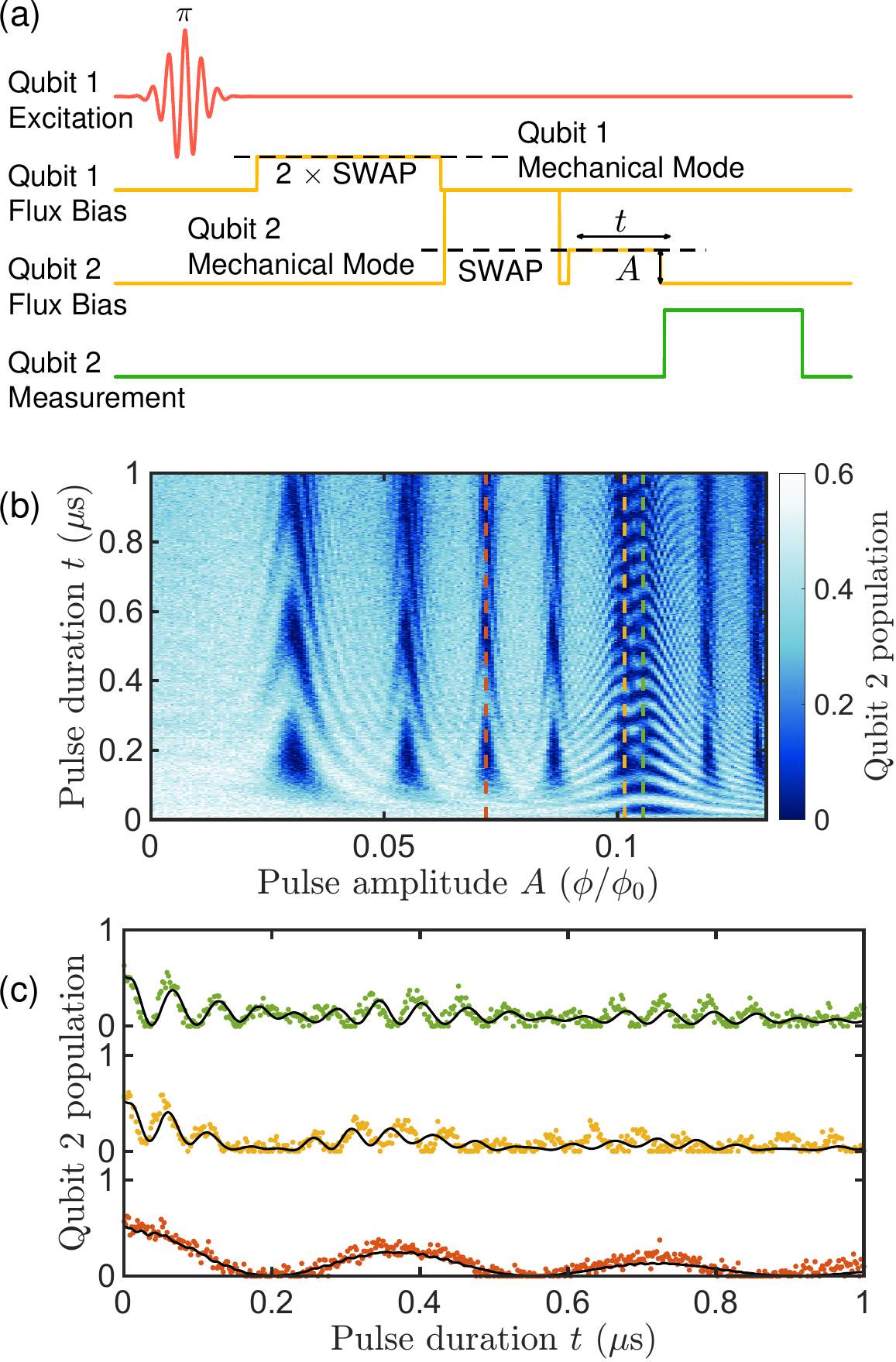}
\caption{\emph{Swapping an excitation throughout the system:} {\bf(a)} The pulse sequence for an experiment where an excitation is swapped between the degrees of freedom of the system. FIRST: Qubit 1 is excited with a $\pi$ pulse, SECOND: the excitation is swapped to a mechanical mode coupled to qubit 1 and back, THIRD: the excitation is swapped to qubit 2, FOURTH: finally a square pulse of variable amplitude and duration is applied to the qubit 2 flux bias so we can see the excitation swapping to the mechanical modes coupled to qubit 2 as well as with qubit 1 again. {\bf(b)} Experimental data using the pulse sequence shown at the top, where qubit 2 is measured after the pulse sequence. The data shows Vacuum Rabi oscillations using an excitation that has traversed the system. {\bf(c)} Line cuts as indicated by the dashed lines in (b) are plotted against the master equation solution.}
\label{fig:SwappedVR}
\end{figure}

It is important to determine whether or not the qubits couple to the same set of acoustic modes. The issue is nontrivial since on one hand, the qubits are in close proximity to each other and share the same HBAR chip, which would point to delocalized acoustic modes. On the other hand, one could argue that the electric field of either qubit should confine the HBAR mode only below its own electrode. We attempted to carry out finite-element simulations, however, full 3-dimensional solution was beyond reach. In 2 dimensions and with a limited overtone number, we saw indications of a delocalized acoustic mode, with the study showing that moving the qubit coupling pad changed the strength of coupling to modes of different lateral profile. Experimentally, the issue cannot immediately be resolved in spectroscopy, since the HBAR spectral lines in figure \ref{fig:overview} are equal within measurement uncertainties, which however is expected based on the geometry. A time-domain experiment was done to confirm that the qubits couple to their individual sets of acoustic modes. This was done by swapping an excitation from qubit 1 to its acoustic mode at 3.788 GHz, and then tuning it away whilst tuning the qubit 2 on resonance with this mode. The experiment found no response and so concluded that the qubits indeed couple to separate modes with any stray coupling being too weak to observe.

Finally, we demonstrate the swapping of an excitation through the degrees of freedom of the system. Figure \ref{fig:SwappedVR} shows the pulse sequence and measured data. The excitation swaps from the 3.7885 GHz HBAR mode coupled to qubit 1 all the way to various HBAR modes coupled to qubit 2. The resulting measurement data is similar to figure \ref{fig:NormalVR} (a) as the last part of the pulse sequence is similar to that experiment, however this excitation has travelled from an acoustic mode coupled to the opposite qubit hence why the initial excited state population is reduced due to decoherence.


Now that we have shown the ability to transfer excitations around the system, we would in principle be able to create an entangled state between arbitrary acoustic modes. However, due to the limited coherence of the system, we were not able to measure this in practice. One needs to measure the entangled modes simultaneously under a series of tomography pulses in order to produce the density matrix of the system (for example see \cite{Wollack2022entangle}). This was not straightforward to do in our system as the acoustic modes are coupled to different qubits, meaning we need to readout the acoustic mode in single-shot to be able to correlate the results. We are limited both by our single-shot readout fidelity $<60\%$, and by not being in the strong dispersive regime which requires acoustic $T_1$ times of 8 $\mu$s at our coupling magnitudes. 

A possible simplification to make is to only measure an entangled state which does not occupy number states higher than $|1\rangle$ so that in this case one can swap the state back to the qubits and measure them. Due to the low  readout fidelity, we have to use an ensemble measurement. There is a tomography pulse scheme to measure the two qubit density matrix using an ensemble measurement \cite{Li2017Ensemble}. This requires an appropriate two-qubit gate as a part of the tomography pulse scheme and in our case this would be an iSWAP pulse. The calibration of this iSWAP pulse was problematic having a fidelity of $55\%$ which was not sufficient to do the two qubit tomography. We estimate that probably higher than $70\%$ gate fidelity  is required to be able to perform the measurement.

In order to improve the fidelity of single and two-qubit gates in the system, one would like the FSR to be larger than the coupling by a factor of at least 20. This is so that if the qubit is in between two modes, it will only interact dispersively. Also the FSR should be larger than inverse pulse widths, so that these are not exciting nearby mechanical modes as well. Longer coherence times for both the qubits and the acoustics are important towards this end. The ideal solution would be the development of a tunable coupler, to be able to selectively couple to modes of interest, which is important for using HBARs in quantum information processing.


In conclusion we have fabricated and measured a sample consisting of two qubits each coupled to an individual set of high overtone bulk acoustic (HBAR) modes as well as to each other. An excitation was swapped from an HBAR mode coupled with one qubit, to an HBAR mode coupled to the other qubit. This demonstrates the possibility to integrate multiple HBAR devices into a superconducting circuit, where complex quantum states could be stored across these devices.

\begin{acknowledgments} We would like to thank Mikael Kervinen for useful discussion. We acknowledge the facilities and technical support of Otaniemi research infrastructure for Micro and Nanotechnologies (OtaNano) that is part of the European Microkelvin Platform. This work was supported by the Academy of Finland (contracts 307757), by the European Research Council (101019712), and by the Wihuri Foundation. We acknowledge funding from the European Union's Horizon 2020 research and innovation program under the QuantERA II Programme (13352189). The work was performed as part of the Academy of Finland Centre of Excellence program (project 336810). 
\end{acknowledgments}


\begin{thebibliography}{20}%
\makeatletter
\providecommand \@ifxundefined [1]{%
 \@ifx{#1\undefined}
}%
\providecommand \@ifnum [1]{%
 \ifnum #1\expandafter \@firstoftwo
 \else \expandafter \@secondoftwo
 \fi
}%
\providecommand \@ifx [1]{%
 \ifx #1\expandafter \@firstoftwo
 \else \expandafter \@secondoftwo
 \fi
}%
\providecommand \natexlab [1]{#1}%
\providecommand \enquote  [1]{``#1''}%
\providecommand \bibnamefont  [1]{#1}%
\providecommand \bibfnamefont [1]{#1}%
\providecommand \citenamefont [1]{#1}%
\providecommand \href@noop [0]{\@secondoftwo}%
\providecommand \href [0]{\begingroup \@sanitize@url \@href}%
\providecommand \@href[1]{\@@startlink{#1}\@@href}%
\providecommand \@@href[1]{\endgroup#1\@@endlink}%
\providecommand \@sanitize@url [0]{\catcode `\\12\catcode `\$12\catcode
  `\&12\catcode `\#12\catcode `\^12\catcode `\_12\catcode `\%12\relax}%
\providecommand \@@startlink[1]{}%
\providecommand \@@endlink[0]{}%
\providecommand \url  [0]{\begingroup\@sanitize@url \@url }%
\providecommand \@url [1]{\endgroup\@href {#1}{\urlprefix }}%
\providecommand \urlprefix  [0]{URL }%
\providecommand \Eprint [0]{\href }%
\providecommand \doibase [0]{http://dx.doi.org/}%
\providecommand \selectlanguage [0]{\@gobble}%
\providecommand \bibinfo  [0]{\@secondoftwo}%
\providecommand \bibfield  [0]{\@secondoftwo}%
\providecommand \translation [1]{[#1]}%
\providecommand \BibitemOpen [0]{}%
\providecommand \bibitemStop [0]{}%
\providecommand \bibitemNoStop [0]{.\EOS\space}%
\providecommand \EOS [0]{\spacefactor3000\relax}%
\providecommand \BibitemShut  [1]{\csname bibitem#1\endcsname}%
\let\auto@bib@innerbib\@empty
\bibitem [{\citenamefont {Clerk}\ \emph {et~al.}(2020)\citenamefont {Clerk},
  \citenamefont {Lehnert}, \citenamefont {Bertet}, \citenamefont {Petta},\ and\
  \citenamefont {Nakamura}}]{Clerk2020hybrid}%
  \BibitemOpen
  \bibfield  {author} {\bibinfo {author} {\bibfnamefont {A.~A.}\ \bibnamefont
  {Clerk}}, \bibinfo {author} {\bibfnamefont {K.~W.}\ \bibnamefont {Lehnert}},
  \bibinfo {author} {\bibfnamefont {P.}~\bibnamefont {Bertet}}, \bibinfo
  {author} {\bibfnamefont {J.~R.}\ \bibnamefont {Petta}}, \ and\ \bibinfo
  {author} {\bibfnamefont {Y.}~\bibnamefont {Nakamura}},\ }\bibfield  {title}
  {\enquote {\bibinfo {title} {Hybrid quantum systems with circuit quantum
  electrodynamics},}\ }\href@noop {} {\bibfield  {journal} {\bibinfo  {journal}
  {Nature Physics}\ }\textbf {\bibinfo {volume} {16}},\ \bibinfo {pages}
  {257--267} (\bibinfo {year} {2020})}\BibitemShut {NoStop}%
\bibitem [{\citenamefont {Regal}\ \emph {et~al.}(2008)\citenamefont {Regal},
  \citenamefont {Teufel},\ and\ \citenamefont {Lehnert}}]{Lehnert2008Nph}%
  \BibitemOpen
  \bibfield  {author} {\bibinfo {author} {\bibfnamefont {C.~A.}\ \bibnamefont
  {Regal}}, \bibinfo {author} {\bibfnamefont {J.~D.}\ \bibnamefont {Teufel}}, \
  and\ \bibinfo {author} {\bibfnamefont {K.~W.}\ \bibnamefont {Lehnert}},\
  }\bibfield  {title} {\enquote {\bibinfo {title} {{Measuring nanomechanical
  motion with a microwave cavity interferometer}},}\ }\href@noop {} {\bibfield
  {journal} {\bibinfo  {journal} {Nature Physics}\ }\textbf {\bibinfo {volume}
  {4}},\ \bibinfo {pages} {555--560} (\bibinfo {year} {2008})}\BibitemShut
  {NoStop}%
\bibitem [{\citenamefont {Teufel}\ \emph {et~al.}(2011)\citenamefont {Teufel},
  \citenamefont {Li}, \citenamefont {Allman}, \citenamefont {Cicak},
  \citenamefont {Sirois}, \citenamefont {Whittaker},\ and\ \citenamefont
  {Simmonds}}]{Teufel2011a}%
  \BibitemOpen
  \bibfield  {author} {\bibinfo {author} {\bibfnamefont {J.~D.}\ \bibnamefont
  {Teufel}}, \bibinfo {author} {\bibfnamefont {Dale}\ \bibnamefont {Li}},
  \bibinfo {author} {\bibfnamefont {M.~S.}\ \bibnamefont {Allman}}, \bibinfo
  {author} {\bibfnamefont {K.}~\bibnamefont {Cicak}}, \bibinfo {author}
  {\bibfnamefont {A.~J.}\ \bibnamefont {Sirois}}, \bibinfo {author}
  {\bibfnamefont {J.~D.}\ \bibnamefont {Whittaker}}, \ and\ \bibinfo {author}
  {\bibfnamefont {R.~W.}\ \bibnamefont {Simmonds}},\ }\bibfield  {title}
  {\enquote {\bibinfo {title} {{Circuit cavity electromechanics in the
  strong-coupling regime}},}\ }\href@noop {} {\bibfield  {journal} {\bibinfo
  {journal} {Nature}\ }\textbf {\bibinfo {volume} {471}},\ \bibinfo {pages}
  {204--208} (\bibinfo {year} {2011})}\BibitemShut {NoStop}%
\bibitem [{\citenamefont {O'Connell}\ \emph {et~al.}(2010)\citenamefont
  {O'Connell}, \citenamefont {Hofheinz}, \citenamefont {Ansmann}, \citenamefont
  {Bialczak}, \citenamefont {Lenander}, \citenamefont {Lucero}, \citenamefont
  {Neeley}, \citenamefont {Sank}, \citenamefont {Wang}, \citenamefont {Weides},
  \citenamefont {Wenner}, \citenamefont {Martinis},\ and\ \citenamefont
  {Cleland}}]{OConnell2010}%
  \BibitemOpen
  \bibfield  {author} {\bibinfo {author} {\bibfnamefont {A.~D.}\ \bibnamefont
  {O'Connell}}, \bibinfo {author} {\bibfnamefont {M.}~\bibnamefont {Hofheinz}},
  \bibinfo {author} {\bibfnamefont {M.}~\bibnamefont {Ansmann}}, \bibinfo
  {author} {\bibfnamefont {Radoslaw~C.}\ \bibnamefont {Bialczak}}, \bibinfo
  {author} {\bibfnamefont {M.}~\bibnamefont {Lenander}}, \bibinfo {author}
  {\bibfnamefont {Erik}\ \bibnamefont {Lucero}}, \bibinfo {author}
  {\bibfnamefont {M.}~\bibnamefont {Neeley}}, \bibinfo {author} {\bibfnamefont
  {D.}~\bibnamefont {Sank}}, \bibinfo {author} {\bibfnamefont {H.}~\bibnamefont
  {Wang}}, \bibinfo {author} {\bibfnamefont {M.}~\bibnamefont {Weides}},
  \bibinfo {author} {\bibfnamefont {J.}~\bibnamefont {Wenner}}, \bibinfo
  {author} {\bibfnamefont {John~M.}\ \bibnamefont {Martinis}}, \ and\ \bibinfo
  {author} {\bibfnamefont {A.~N.}\ \bibnamefont {Cleland}},\ }\bibfield
  {title} {\enquote {\bibinfo {title} {{Quantum ground state and single-phonon
  control of a mechanical resonator}},}\ }\href@noop {} {\bibfield  {journal}
  {\bibinfo  {journal} {Nature}\ }\textbf {\bibinfo {volume} {464}},\ \bibinfo
  {pages} {697--703} (\bibinfo {year} {2010})}\BibitemShut {NoStop}%
\bibitem [{\citenamefont {Arrangoiz-Arriola}\ \emph {et~al.}(2019)\citenamefont
  {Arrangoiz-Arriola}, \citenamefont {Wollack}, \citenamefont {Wang},
  \citenamefont {Pechal}, \citenamefont {Jiang}, \citenamefont {McKenna},
  \citenamefont {Witmer}, \citenamefont {Van~Laer},\ and\ \citenamefont
  {Safavi-Naeini}}]{Safavi2019Fock}%
  \BibitemOpen
  \bibfield  {author} {\bibinfo {author} {\bibfnamefont {Patricio}\
  \bibnamefont {Arrangoiz-Arriola}}, \bibinfo {author} {\bibfnamefont
  {E.~Alex}\ \bibnamefont {Wollack}}, \bibinfo {author} {\bibfnamefont
  {Zhaoyou}\ \bibnamefont {Wang}}, \bibinfo {author} {\bibfnamefont {Marek}\
  \bibnamefont {Pechal}}, \bibinfo {author} {\bibfnamefont {Wentao}\
  \bibnamefont {Jiang}}, \bibinfo {author} {\bibfnamefont {Timothy~P.}\
  \bibnamefont {McKenna}}, \bibinfo {author} {\bibfnamefont {Jeremy~D.}\
  \bibnamefont {Witmer}}, \bibinfo {author} {\bibfnamefont {Rapha{\"e}l}\
  \bibnamefont {Van~Laer}}, \ and\ \bibinfo {author} {\bibfnamefont {Amir~H.}\
  \bibnamefont {Safavi-Naeini}},\ }\bibfield  {title} {\enquote {\bibinfo
  {title} {Resolving the energy levels of a nanomechanical oscillator},}\
  }\href@noop {} {\bibfield  {journal} {\bibinfo  {journal} {Nature}\ }\textbf
  {\bibinfo {volume} {571}},\ \bibinfo {pages} {537--540} (\bibinfo {year}
  {2019})}\BibitemShut {NoStop}%
\bibitem [{\citenamefont {Chu}\ \emph {et~al.}(2017)\citenamefont {Chu},
  \citenamefont {Kharel}, \citenamefont {Renninger}, \citenamefont {Burkhart},
  \citenamefont {Frunzio}, \citenamefont {Rakich},\ and\ \citenamefont
  {Schoelkopf}}]{SchoelkopfHBAR2017}%
  \BibitemOpen
  \bibfield  {author} {\bibinfo {author} {\bibfnamefont {Yiwen}\ \bibnamefont
  {Chu}}, \bibinfo {author} {\bibfnamefont {Prashanta}\ \bibnamefont {Kharel}},
  \bibinfo {author} {\bibfnamefont {William~H.}\ \bibnamefont {Renninger}},
  \bibinfo {author} {\bibfnamefont {Luke~D.}\ \bibnamefont {Burkhart}},
  \bibinfo {author} {\bibfnamefont {Luigi}\ \bibnamefont {Frunzio}}, \bibinfo
  {author} {\bibfnamefont {Peter~T.}\ \bibnamefont {Rakich}}, \ and\ \bibinfo
  {author} {\bibfnamefont {Robert~J.}\ \bibnamefont {Schoelkopf}},\ }\bibfield
  {title} {\enquote {\bibinfo {title} {Quantum acoustics with superconducting
  qubits},}\ }\href@noop {} {\bibfield  {journal} {\bibinfo  {journal}
  {Science}\ }\textbf {\bibinfo {volume} {358}},\ \bibinfo {pages} {199--202}
  (\bibinfo {year} {2017})}\BibitemShut {NoStop}%
\bibitem [{\citenamefont {Kervinen}\ \emph {et~al.}(2018)\citenamefont
  {Kervinen}, \citenamefont {Rissanen},\ and\ \citenamefont
  {Sillanp\"a\"a}}]{kervinen_interfacing_2018}%
  \BibitemOpen
  \bibfield  {author} {\bibinfo {author} {\bibfnamefont {Mikael}\ \bibnamefont
  {Kervinen}}, \bibinfo {author} {\bibfnamefont {Ilkka}\ \bibnamefont
  {Rissanen}}, \ and\ \bibinfo {author} {\bibfnamefont {Mika}\ \bibnamefont
  {Sillanp\"a\"a}},\ }\bibfield  {title} {\enquote {\bibinfo {title}
  {{Interfacing planar superconducting qubits with high overtone bulk acoustic
  phonons}},}\ }\href@noop {} {\bibfield  {journal} {\bibinfo  {journal}
  {Physical Review B}\ }\textbf {\bibinfo {volume} {97}},\ \bibinfo {pages}
  {205443} (\bibinfo {year} {2018})}\BibitemShut {NoStop}%
\bibitem [{\citenamefont {Gustafsson}\ \emph {et~al.}(2014)\citenamefont
  {Gustafsson}, \citenamefont {Aref}, \citenamefont {Kockum}, \citenamefont
  {Ekstr\"om}, \citenamefont {Johansson},\ and\ \citenamefont
  {Delsing}}]{Delsing2014}%
  \BibitemOpen
  \bibfield  {author} {\bibinfo {author} {\bibfnamefont {Martin~V.}\
  \bibnamefont {Gustafsson}}, \bibinfo {author} {\bibfnamefont {Thomas}\
  \bibnamefont {Aref}}, \bibinfo {author} {\bibfnamefont {Anton~Frisk}\
  \bibnamefont {Kockum}}, \bibinfo {author} {\bibfnamefont {Maria~K.}\
  \bibnamefont {Ekstr\"om}}, \bibinfo {author} {\bibfnamefont {G\"oran}\
  \bibnamefont {Johansson}}, \ and\ \bibinfo {author} {\bibfnamefont {Per}\
  \bibnamefont {Delsing}},\ }\bibfield  {title} {\enquote {\bibinfo {title}
  {Propagating phonons coupled to an artificial atom},}\ }\href@noop {}
  {\bibfield  {journal} {\bibinfo  {journal} {Science}\ }\textbf {\bibinfo
  {volume} {346}},\ \bibinfo {pages} {207--211} (\bibinfo {year}
  {2014})}\BibitemShut {NoStop}%
\bibitem [{\citenamefont {Noguchi}\ \emph {et~al.}(2017)\citenamefont
  {Noguchi}, \citenamefont {Yamazaki}, \citenamefont {Tabuchi},\ and\
  \citenamefont {Nakamura}}]{Nakamura2017}%
  \BibitemOpen
  \bibfield  {author} {\bibinfo {author} {\bibfnamefont {Atsushi}\ \bibnamefont
  {Noguchi}}, \bibinfo {author} {\bibfnamefont {Rekishu}\ \bibnamefont
  {Yamazaki}}, \bibinfo {author} {\bibfnamefont {Yutaka}\ \bibnamefont
  {Tabuchi}}, \ and\ \bibinfo {author} {\bibfnamefont {Yasunobu}\ \bibnamefont
  {Nakamura}},\ }\bibfield  {title} {\enquote {\bibinfo {title} {Qubit-assisted
  transduction for a detection of surface acoustic waves near the quantum
  limit},}\ }\href@noop {} {\bibfield  {journal} {\bibinfo  {journal} {Phys.
  Rev. Lett.}\ }\textbf {\bibinfo {volume} {119}},\ \bibinfo {pages} {180505}
  (\bibinfo {year} {2017})}\BibitemShut {NoStop}%
\bibitem [{\citenamefont {Moores}\ \emph {et~al.}(2018)\citenamefont {Moores},
  \citenamefont {Sletten}, \citenamefont {Viennot},\ and\ \citenamefont
  {Lehnert}}]{moores_cavity_2018}%
  \BibitemOpen
  \bibfield  {author} {\bibinfo {author} {\bibfnamefont {Bradley~A.}\
  \bibnamefont {Moores}}, \bibinfo {author} {\bibfnamefont {Lucas~R.}\
  \bibnamefont {Sletten}}, \bibinfo {author} {\bibfnamefont {Jeremie~J.}\
  \bibnamefont {Viennot}}, \ and\ \bibinfo {author} {\bibfnamefont {K.~W.}\
  \bibnamefont {Lehnert}},\ }\bibfield  {title} {\enquote {\bibinfo {title}
  {Cavity {Quantum} {Acoustic} {Device} in the {Multimode} {Strong} {Coupling}
  {Regime}},}\ }\href@noop {} {\bibfield  {journal} {\bibinfo  {journal}
  {Physical Review Letters}\ }\textbf {\bibinfo {volume} {120}},\ \bibinfo
  {pages} {227701} (\bibinfo {year} {2018})}\BibitemShut {NoStop}%
\bibitem [{\citenamefont {Bienfait}\ \emph {et~al.}(2019)\citenamefont
  {Bienfait}, \citenamefont {Satzinger}, \citenamefont {Zhong}, \citenamefont
  {Chang}, \citenamefont {Chou}, \citenamefont {Conner}, \citenamefont {Dumur},
  \citenamefont {Grebel}, \citenamefont {Peairs}, \citenamefont {Povey},\ and\
  \citenamefont {Cleland}}]{Cleland2019PhEntangl}%
  \BibitemOpen
  \bibfield  {author} {\bibinfo {author} {\bibfnamefont {A.}~\bibnamefont
  {Bienfait}}, \bibinfo {author} {\bibfnamefont {K.~J.}\ \bibnamefont
  {Satzinger}}, \bibinfo {author} {\bibfnamefont {Y.~P.}\ \bibnamefont
  {Zhong}}, \bibinfo {author} {\bibfnamefont {H.-S.}\ \bibnamefont {Chang}},
  \bibinfo {author} {\bibfnamefont {M.-H.}\ \bibnamefont {Chou}}, \bibinfo
  {author} {\bibfnamefont {C.~R.}\ \bibnamefont {Conner}}, \bibinfo {author}
  {\bibfnamefont {{\'E}.}~\bibnamefont {Dumur}}, \bibinfo {author}
  {\bibfnamefont {J.}~\bibnamefont {Grebel}}, \bibinfo {author} {\bibfnamefont
  {G.~A.}\ \bibnamefont {Peairs}}, \bibinfo {author} {\bibfnamefont {R.~G.}\
  \bibnamefont {Povey}}, \ and\ \bibinfo {author} {\bibfnamefont {A.~N.}\
  \bibnamefont {Cleland}},\ }\bibfield  {title} {\enquote {\bibinfo {title}
  {Phonon-mediated quantum state transfer and remote qubit entanglement},}\
  }\href@noop {} {\bibfield  {journal} {\bibinfo  {journal} {Science}\ }\textbf
  {\bibinfo {volume} {364}},\ \bibinfo {pages} {368--371} (\bibinfo {year}
  {2019})}\BibitemShut {NoStop}%
\bibitem [{\citenamefont {Gokhale}\ \emph {et~al.}(2020)\citenamefont
  {Gokhale}, \citenamefont {Downey}, \citenamefont {Katzer}, \citenamefont
  {Nepal}, \citenamefont {Lang}, \citenamefont {Stroud},\ and\ \citenamefont
  {Meyer}}]{gokhale_epitaxial_2020}%
  \BibitemOpen
  \bibfield  {author} {\bibinfo {author} {\bibfnamefont {Vikrant~J.}\
  \bibnamefont {Gokhale}}, \bibinfo {author} {\bibfnamefont {Brian~P.}\
  \bibnamefont {Downey}}, \bibinfo {author} {\bibfnamefont {D.~Scott}\
  \bibnamefont {Katzer}}, \bibinfo {author} {\bibfnamefont {Neeraj}\
  \bibnamefont {Nepal}}, \bibinfo {author} {\bibfnamefont {Andrew~C.}\
  \bibnamefont {Lang}}, \bibinfo {author} {\bibfnamefont {Rhonda~M.}\
  \bibnamefont {Stroud}}, \ and\ \bibinfo {author} {\bibfnamefont {David~J.}\
  \bibnamefont {Meyer}},\ }\bibfield  {title} {\enquote {\bibinfo {title}
  {Epitaxial bulk acoustic wave resonators as highly coherent multi-phonon
  sources for quantum acoustodynamics},}\ }\href@noop {} {\bibfield  {journal}
  {\bibinfo  {journal} {Nature Communications}\ }\textbf {\bibinfo {volume}
  {11}},\ \bibinfo {pages} {2314} (\bibinfo {year} {2020})}\BibitemShut
  {NoStop}%
\bibitem [{\citenamefont {Chu}\ \emph {et~al.}(2018)\citenamefont {Chu},
  \citenamefont {Kharel}, \citenamefont {Yoon}, \citenamefont {Frunzio},
  \citenamefont {Rakich},\ and\ \citenamefont
  {Schoelkopf}}]{chu_creation_2018}%
  \BibitemOpen
  \bibfield  {author} {\bibinfo {author} {\bibfnamefont {Yiwen}\ \bibnamefont
  {Chu}}, \bibinfo {author} {\bibfnamefont {Prashanta}\ \bibnamefont {Kharel}},
  \bibinfo {author} {\bibfnamefont {Taekwan}\ \bibnamefont {Yoon}}, \bibinfo
  {author} {\bibfnamefont {Luigi}\ \bibnamefont {Frunzio}}, \bibinfo {author}
  {\bibfnamefont {Peter~T.}\ \bibnamefont {Rakich}}, \ and\ \bibinfo {author}
  {\bibfnamefont {Robert~J.}\ \bibnamefont {Schoelkopf}},\ }\bibfield  {title}
  {\enquote {\bibinfo {title} {{Creation and control of multi-phonon Fock
  states in a bulk acoustic-wave resonator}},}\ }\href@noop {} {\bibfield
  {journal} {\bibinfo  {journal} {Nature}\ }\textbf {\bibinfo {volume} {563}},\
  \bibinfo {pages} {666--670} (\bibinfo {year} {2018})}\BibitemShut {NoStop}%
\bibitem [{\citenamefont {Kervinen}\ \emph {et~al.}(2019)\citenamefont
  {Kervinen}, \citenamefont {Ram\'{\i}rez-Mu\~noz}, \citenamefont {V\"alimaa},\
  and\ \citenamefont {Sillanp\"a\"a}}]{kervinen2019landau}%
  \BibitemOpen
  \bibfield  {author} {\bibinfo {author} {\bibfnamefont {Mikael}\ \bibnamefont
  {Kervinen}}, \bibinfo {author} {\bibfnamefont {Jhon~E.}\ \bibnamefont
  {Ram\'{\i}rez-Mu\~noz}}, \bibinfo {author} {\bibfnamefont {Alpo}\
  \bibnamefont {V\"alimaa}}, \ and\ \bibinfo {author} {\bibfnamefont {Mika~A.}\
  \bibnamefont {Sillanp\"a\"a}},\ }\bibfield  {title} {\enquote {\bibinfo
  {title} {{Landau-Zener-St\"uckelberg Interference in a Multimode
  Electromechanical System in the Quantum Regime}},}\ }\href@noop {} {\bibfield
   {journal} {\bibinfo  {journal} {Phys. Rev. Lett.}\ }\textbf {\bibinfo
  {volume} {123}},\ \bibinfo {pages} {240401} (\bibinfo {year}
  {2019})}\BibitemShut {NoStop}%
\bibitem [{\citenamefont {Wollack}\ \emph {et~al.}(2022)\citenamefont
  {Wollack}, \citenamefont {Cleland}, \citenamefont {Gruenke}, \citenamefont
  {Wang}, \citenamefont {Arrangoiz-Arriola},\ and\ \citenamefont
  {Safavi-Naeini}}]{Wollack2022entangle}%
  \BibitemOpen
  \bibfield  {author} {\bibinfo {author} {\bibfnamefont {E.~Alex}\ \bibnamefont
  {Wollack}}, \bibinfo {author} {\bibfnamefont {Agnetta~Y.}\ \bibnamefont
  {Cleland}}, \bibinfo {author} {\bibfnamefont {Rachel~G.}\ \bibnamefont
  {Gruenke}}, \bibinfo {author} {\bibfnamefont {Zhaoyou}\ \bibnamefont {Wang}},
  \bibinfo {author} {\bibfnamefont {Patricio}\ \bibnamefont
  {Arrangoiz-Arriola}}, \ and\ \bibinfo {author} {\bibfnamefont {Amir~H.}\
  \bibnamefont {Safavi-Naeini}},\ }\bibfield  {title} {\enquote {\bibinfo
  {title} {Quantum state preparation and tomography of entangled mechanical
  resonators},}\ }\href@noop {} {\bibfield  {journal} {\bibinfo  {journal}
  {Nature}\ }\textbf {\bibinfo {volume} {604}},\ \bibinfo {pages} {463--467}
  (\bibinfo {year} {2022})}\BibitemShut {NoStop}%
\bibitem [{\citenamefont {Kervinen}\ \emph {et~al.}(2020)\citenamefont
  {Kervinen}, \citenamefont {V\"alimaa}, \citenamefont {Ram\'{\i}rez-Mu\~noz},\
  and\ \citenamefont {Sillanp\"a\"a}}]{Kervinen2020}%
  \BibitemOpen
  \bibfield  {author} {\bibinfo {author} {\bibfnamefont {Mikael}\ \bibnamefont
  {Kervinen}}, \bibinfo {author} {\bibfnamefont {Alpo}\ \bibnamefont
  {V\"alimaa}}, \bibinfo {author} {\bibfnamefont {Jhon~E.}\ \bibnamefont
  {Ram\'{\i}rez-Mu\~noz}}, \ and\ \bibinfo {author} {\bibfnamefont {Mika~A.}\
  \bibnamefont {Sillanp\"a\"a}},\ }\bibfield  {title} {\enquote {\bibinfo
  {title} {Sideband control of a multimode quantum bulk acoustic system},}\
  }\href@noop {} {\bibfield  {journal} {\bibinfo  {journal} {Phys. Rev.
  Applied}\ }\textbf {\bibinfo {volume} {14}},\ \bibinfo {pages} {054023}
  (\bibinfo {year} {2020})}\BibitemShut {NoStop}%
\bibitem [{\citenamefont {von L{\"u}pke}\ \emph {et~al.}(2022)\citenamefont
  {von L{\"u}pke}, \citenamefont {Yang}, \citenamefont {Bild}, \citenamefont
  {Michaud}, \citenamefont {Fadel},\ and\ \citenamefont {Chu}}]{Lupke2022}%
  \BibitemOpen
  \bibfield  {author} {\bibinfo {author} {\bibfnamefont {Uwe}\ \bibnamefont
  {von L{\"u}pke}}, \bibinfo {author} {\bibfnamefont {Yu}~\bibnamefont {Yang}},
  \bibinfo {author} {\bibfnamefont {Marius}\ \bibnamefont {Bild}}, \bibinfo
  {author} {\bibfnamefont {Laurent}\ \bibnamefont {Michaud}}, \bibinfo {author}
  {\bibfnamefont {Matteo}\ \bibnamefont {Fadel}}, \ and\ \bibinfo {author}
  {\bibfnamefont {Yiwen}\ \bibnamefont {Chu}},\ }\bibfield  {title} {\enquote
  {\bibinfo {title} {Parity measurement in the strong dispersive regime of
  circuit quantum acoustodynamics},}\ }\href@noop {} {\bibfield  {journal}
  {\bibinfo  {journal} {Nature Physics}\ }\textbf {\bibinfo {volume} {18}},\
  \bibinfo {pages} {794--799} (\bibinfo {year} {2022})}\BibitemShut {NoStop}%
\bibitem [{\citenamefont {Kwon}\ \emph {et~al.}(2021)\citenamefont {Kwon},
  \citenamefont {Tomonaga}, \citenamefont {Lakshmi~Bhai}, \citenamefont
  {Devitt},\ and\ \citenamefont {Tsai}}]{Kwon2021GateBased}%
  \BibitemOpen
  \bibfield  {author} {\bibinfo {author} {\bibfnamefont {Sangil}\ \bibnamefont
  {Kwon}}, \bibinfo {author} {\bibfnamefont {Akiyoshi}\ \bibnamefont
  {Tomonaga}}, \bibinfo {author} {\bibfnamefont {Gopika}\ \bibnamefont
  {Lakshmi~Bhai}}, \bibinfo {author} {\bibfnamefont {Simon~J.}\ \bibnamefont
  {Devitt}}, \ and\ \bibinfo {author} {\bibfnamefont {Jaw-Shen}\ \bibnamefont
  {Tsai}},\ }\bibfield  {title} {\enquote {\bibinfo {title} {Gate-based
  superconducting quantum computing},}\ }\href@noop {} {\bibfield  {journal}
  {\bibinfo  {journal} {Journal of Applied Physics}\ }\textbf {\bibinfo
  {volume} {129}},\ \bibinfo {pages} {041102} (\bibinfo {year}
  {2021})}\BibitemShut {NoStop}%
\bibitem [{\citenamefont {Rol}\ \emph {et~al.}(2020)\citenamefont {Rol},
  \citenamefont {Ciorciaro}, \citenamefont {Malinowski}, \citenamefont
  {Tarasinski}, \citenamefont {Sagastizabal}, \citenamefont {Bultink},
  \citenamefont {Salathe}, \citenamefont {Haandbaek}, \citenamefont {Sedivy},\
  and\ \citenamefont {DiCarlo}}]{Rol2020PulseDistorsion}%
  \BibitemOpen
  \bibfield  {author} {\bibinfo {author} {\bibfnamefont {M.~A.}\ \bibnamefont
  {Rol}}, \bibinfo {author} {\bibfnamefont {L.}~\bibnamefont {Ciorciaro}},
  \bibinfo {author} {\bibfnamefont {F.~K.}\ \bibnamefont {Malinowski}},
  \bibinfo {author} {\bibfnamefont {B.~M.}\ \bibnamefont {Tarasinski}},
  \bibinfo {author} {\bibfnamefont {R.~E.}\ \bibnamefont {Sagastizabal}},
  \bibinfo {author} {\bibfnamefont {C.~C.}\ \bibnamefont {Bultink}}, \bibinfo
  {author} {\bibfnamefont {Y.}~\bibnamefont {Salathe}}, \bibinfo {author}
  {\bibfnamefont {N.}~\bibnamefont {Haandbaek}}, \bibinfo {author}
  {\bibfnamefont {J.}~\bibnamefont {Sedivy}}, \ and\ \bibinfo {author}
  {\bibfnamefont {L.}~\bibnamefont {DiCarlo}},\ }\bibfield  {title} {\enquote
  {\bibinfo {title} {{Time-domain characterization and correction of on-chip
  distortion of control pulses in a quantum processor}},}\ }\href@noop {}
  {\bibfield  {journal} {\bibinfo  {journal} {Applied Physics Letters}\
  }\textbf {\bibinfo {volume} {116}},\ \bibinfo {pages} {054001} (\bibinfo
  {year} {2020})}\BibitemShut {NoStop}%
\bibitem [{\citenamefont {Li}\ \emph {et~al.}(2017)\citenamefont {Li},
  \citenamefont {Xue}, \citenamefont {Tan}, \citenamefont {Liu}, \citenamefont
  {Dai}, \citenamefont {Zhang}, \citenamefont {Yu},\ and\ \citenamefont
  {Yu}}]{Li2017Ensemble}%
  \BibitemOpen
  \bibfield  {author} {\bibinfo {author} {\bibfnamefont {Mengmeng}\
  \bibnamefont {Li}}, \bibinfo {author} {\bibfnamefont {Guangming}\
  \bibnamefont {Xue}}, \bibinfo {author} {\bibfnamefont {Xinsheng}\
  \bibnamefont {Tan}}, \bibinfo {author} {\bibfnamefont {Qiang}\ \bibnamefont
  {Liu}}, \bibinfo {author} {\bibfnamefont {Kunzhe}\ \bibnamefont {Dai}},
  \bibinfo {author} {\bibfnamefont {Ke}~\bibnamefont {Zhang}}, \bibinfo
  {author} {\bibfnamefont {Haifeng}\ \bibnamefont {Yu}}, \ and\ \bibinfo
  {author} {\bibfnamefont {Yang}\ \bibnamefont {Yu}},\ }\bibfield  {title}
  {\enquote {\bibinfo {title} {{Two-qubit state tomography with ensemble
  average in coupled superconducting qubits}},}\ }\href@noop {} {\bibfield
  {journal} {\bibinfo  {journal} {Applied Physics Letters}\ }\textbf {\bibinfo
  {volume} {110}},\ \bibinfo {pages} {132602} (\bibinfo {year}
  {2017})}\BibitemShut {NoStop}%
\end{thebibliography}

%

\end{document}